\documentclass{kluwer}

\usepackage{graphicx}

\usepackage{times}

\begin{document}

\begin{article}

\begin{opening}

\title{Dynamical Evolution of Planets in Disks}
\runningtitle{Evolution of Planets in Disks}
\subtitle{Planets in resonant Orbits}

\author{W. \surname{Kley}\email{wilhelm.kley@uni-tuebingen.de}}
\runningauthor{W. Kley}
\institute{Astronomie und Astrophysik, Abt. Computational Physics,
Universit\"at T\"ubingen, D-72076 T\"ubingen, Germany}

\received{\ldots} \accepted{\ldots}

\begin{ao}
W. Kley, Astronomie und Astrophysik, Abt. Computational Physics,
Universit\"at T\"ubingen, D-72076 T\"ubingen, Germany,
wilhelm.kley@uni-tuebingen.de
\end{ao}
\begin{abstract}
We study the evolution of a system consisting of two protoplanets
still embedded in a protoplanetary disk.
Results of two different numerical approaches are presented. In the first
kind of model the motion of the disk material is followed by fully
viscous hydrodynamical simulations, and the planetary motion is
determined by N-body calculations including exactly the gravitational
potential from the disk material.
In the second kind we only solve the N-body part and add additional
analytically given forces which model the effect of the torques of the
disk. This type of modeling is of course orders of magnitudes faster
than the full hydro-model.
Another advantage of this two-fold approach is the possibility
of adjusting the otherwise unknown parameters of the simplified model.

The results give very good agreement between the methods. Using two
different initial setups for the planets and disk, we obtain 
in the first case a resonant trapping into the 3:1 resonance, and
in the second case a trapping into the 2:1 resonance. 
Resonant capture leads to a rise in the eccentricity and to an
alignment of the of the spatial orientation of orbits.
The characteristics of the numerical results agree very favorably with those
of 3 observed planetary systems (GJ~876, HD 82943, and 55~Cnc)
known to be in mean motion resonances.
\end{abstract}
\end{opening}
\section{Introduction}
\label{sec:introduction}
Since the first discovery in 1995, during the last years the number
of detected extrasolar planets
orbiting solar type stars has risen up to about 100 
(see
eg. {\tt {http://www.obspm.fr/encycl/encycl.html}} by J.~Schneider
for an always up to date list). 
It has been found that among those there are 11
systems with 2 or more planets.
With further observations to come, this
number may still increase, as for some systems trends in the
radial velocity curve have been found.
Among these multiple planet extrasolar systems there are now 3 confirmed cases,
GJ~876 \cite{2001ApJ...556..296M},
HD 82943 (the {\it Coralie Planet Search Programme},
ESO Press Release 07/01), 55 Cnc \cite{2002ApJ...581.1375M}
where the planets orbit their central star in a low order 
{\it mean motion resonance},
where the orbital periods have nearly exactly the ratios 2:1 or 3:1.
The parameter of these planetary systems
are displayed below in Table~\ref{tab:system}.
This implies that about 1/4 of planetary systems, or even more,
may be in resonance,
a fraction which is even higher if secular resonances,
as for example present in $\upsilon$ And \cite{2002ApJ...576..473C},
are also taken into account.

The formation of such resonant planetary systems can be understood
by considering the joint evolution of proto-planets together with in the
protoplanetary disk from which they formed.
By local linear analysis it was shown that
the gravitational interaction of a protoplanet with
the disk may lead to torques resulting in the migration of a
planet
\cite{1980ApJ...241..425G,1986ApJ...309..846L,1997Icar..126..261W,2002ApJ...565.1257T}.
Additionally, for planetary masses of around one Jupiter mass,
gap formation as result of angular momentum
transfer between the (viscous) disk and the planet was
considered \cite{1993prpl.conf..749L}.
Fully non-linear hydrodynamical calculations
\cite{1999MNRAS.303..696K,1999ApJ...514..344B,1999ApJ...526.1001L,2000MNRAS.318...18N}
for Jupiter sized planets confirmed the expectations and showed
clearly that this interaction leads to:
{\it i)} the formation of spiral shocks waves in the disk, whose tightness
depends on the sound-speed in the disk,
{\it ii)} an annular gap, whose width is determined by the equilibrium
between gap opening tidal torques and gap closing viscous and pressure
forces.
{\it iii)} an inward migration on a timescale of $10^5$ yrs for typical
disk parameter in particular disk masses corresponding to that of the
minimum mass solar nebula,
{\it iv)} a possible mass growth after gap formation up to about
10 $M_{Jup}$ when finally gravitational torques overwhelm, and
finally 
{\it v)} a prograde rotation of the planet.

Recently, these single planet calculations were extended to
calculations with multiple planets.
Those have shown already
\cite{2000MNRAS.313L..47K,2000ApJ...540.1091B,2001A&A...374.1092S,2002MNRAS.333L..26N}
that during the early evolution
of protoplanetary systems, when the planets are still embedded in the disk,
different migration speeds may lead an approach of the
planets and eventually to resonant capture.
Pure N-body calculations including additional damping terms
to model disk-planet interaction effects have been calculated
fo example by \cite{2001A&A...374.1092S,2002ApJ...567..596L,2002ApJ...565..608M}.

Here we present new numerical calculations treating the
evolution of a pair of two embedded planets in disks.
We consider both, fully
hydrodynamic and simplified N-body calculations to model
the evolution. In the first approach, the motion of the disk is followed
by solving the full time dependent Navier-Stokes equations simultaneously
with the motion of the planets. Here the motion of the planet
is determined by the gravitational potential of the other planet,
the star, and that of the disk.
In the latter approach we take a simplified
assumption and perform 3-body (star and 2 planets) calculations
augmented by additional (damping) forces which take the gravity
of the disk approximately into account.
These two approaches allow a direct comparison of the methods,
and will enable us to determine in detail the damping constants
required for the simpler (and much faster) second model.
\section{The Observations}
\label{sec:obs}
The basic orbital parameter of the 3 systems in mean motion resonance
are stated in Table~\ref{tab:system}. Two of them,
GJ~876 and HD~82943 are in a nearly exact 2:1 resonance.
In both cases we note that the outer planet is the more massive one
by a factor of about two (HD~82943), and more than three (GJ~876).
The eccentricity of the inner (less massive) planet is larger than
that of the outer one. For the system GJ~876 the alignment of the orbits
is such that the two periastrae are pointing in nearly the same
direction. The values quoted in Table~\ref{tab:system} are based on the
dynamical orbit calcuations to match the combined Keck+Lick data
\cite{2002ApJ...567..596L,2001ApJ...551L.109L}.
For the system HD~82943 these data have not been
clearly identified, due to the much longer orbital periods,
but do not seem to very different from each other.
The last system, 55~Cnc, is actually a triple system. Here the inner
two planets orbit the star very closely and are in a 3:1
resonance, while the additional, more massive planet orbits at a
distance of several AU \cite{2002ApJ...581.1375M}. For other
longer period systems where only trends in the radial
velocity curve have been observed further observations may
reveal even more systems in a resonant configuration.
\begin{table}
\caption{
The orbital parameter of the 3 systems known to be in a mean
motion resonance. $P$ denotes the orbital period,  $M\sin i$ the
mass of the planets, $a$ the semi-major axis, $e$ the eccentricity,
$\omega$ the longitude of periastron, and $M_*$ the mass of the central
star. It should be noted, that the orbital elements for the
planets may vary on secular time scales. Thus in principle one should
always state the epoch corresponding to these osculating elements.
The values quoted for
GJ~876 are based such on a dynamical fit to the data 
\protect\cite{2002ApJ...567..596L,2001ApJ...551L.109L}, where the
planetary mass refers to the real masses assuming $\sin i = 0.78$.
See also text.
}
\label{tab:system}
\begin{tabular}{ll|l|l|l|l|l|ll} \hline
  &  Nr  &   Per  & $M\sin i$  &  a        &   $e$  &  $\omega$ & $M_*$ & \\
  &    &   [d]  &  $M_{Jup}$  &  [AU]   &        &   [deg] &  $M_\odot$  & \\
 \hline
 \multicolumn{5}{c} { {GJ~876}  (2:1) }\\
  & c   &   30.569 &   0.766        &  0.13   &   0.24  &  159  &  0.32 &  \\
  & b   &   60.128 &   2.403        &  0.21  &   0.04  &  163  &   &  \\
 \hline
 \multicolumn{5}{c} { HD~82943 (2:1) }\\
  & b   &  221.6  &  0.88          &  0.73   &   0.54 &   138 &  1.05 & \\
  & c   &  444.6  &  1.63          &  1.16   &   0.41 &    96 &   & \\
 \hline
 \multicolumn{5}{c}{55~Cnc (3:1) }\\
  & b   &  14.65 &  0.84          &  0.11   &   0.02 &   99  & 0.95 & \\
  & c   &  44.26 &   0.21         &  0.24   &   0.34 &    61 &  & \\
  & d   & 5360   &  4.05          &  5.9    &   0.16 &   201 &  &  \\
 \hline
\end{tabular}
\end{table}
\section{The Models}
\label{sec:model}
It is our goal to determine the evolution of protoplanets
still embedded in their disks. To this purpose we employ two
different methods which supplement each other. Firstly, a fully
time-dependent hydrodynamical model for the joint evolution
of the planets {\it and} disk is presented.
Because the evolutionary time scale may cover several thousands of orbits
these computations require sometimes millions of time-steps, which
translates into  an effective computational times of up to several
weeks.

Because often the main interest focuses only on the orbital
evolution of the planet and not so much on the hydrodynamics of the
disk, we perform additional simplified 3-body computations. Here
the gravitational forces are augmented by additional damping terms
designed in such a way as to incorporate in a simplified way the
gravitational influence of the disk.
Through a direct comparison with the hydrodynamical model
it is then possible to infer directly the necessary damping forces.
\subsection{Full Hydrodynamics}
\label{subsec:model-fullhydro}
The first set of coupled hydrodynamical-N-body
models are calculated in the same manner as described in
detail in \cite{1998A&A...338L..37K}, \cite{1999MNRAS.303..696K}
for single planets and in \cite{2000MNRAS.313L..47K}
for multiple planets, and the reader is referred to those papers
for more details on the computational aspect of the simulations
the computations.
Other similar models, following explicitly the motion of single and
multiple planets in disks, have been presented by 
\cite{2000MNRAS.318...18N,2000ApJ...540.1091B,2001A&A...374.1092S}.
During the evolution material is taken out from the centers
of the Roche-lobes of the two planets, which is monitored and
assumed to have been accreted onto the two planets.
We present two runs: one (model B) where the mass is added to the planet,
and another one (model A) where 
this mass is not added to the dynamical mass of the planets, i.e they
always keep their initial mass. 
They are allowed to migrate (change their semi-major axis)
through the disk according to the gravitational
torques exerted on them. 
This assumption of constant planet mass throughout the
computation is well justified, as the
migration rate depends, at least for type II drift,
only weakly on the mass of the planet
\cite{2000MNRAS.318...18N}.
\begin{table}
\caption{
Planetary and disk parameter of the models. The mass of the Planet is
given in Jupiter masses ($M_{Jup} = 10^{-3} M_{\odot}$),
the viscosity in dimensionless units, 
the disk mass located between $r_{min}$ and $r_{max}$in solar masses,
and the minimum and maximum radii in AU.
}
\label{tab:modpar}
\begin{tabular}{lllllll}
 \hline
Model  &   Mass1  &  Mass2  &    Viscosity  & $M_{disk}$  
  & $r_{min}$ &  $r_{max}$\\
 \hline
  A    &    3    &    5    &  $\alpha = 10^{-2}$ 
  &    0.01  &   1 &  30  \\
  B    &   1 (Var)   &    1 (Var)   &    $\nu =$ const. 
  &    0.01  &   1 &  20  \\
 \hline
\end{tabular}
\end{table}
The initial hydrodynamic structure of the disk, which extends radially
from $r_{min}$ to $r_{max}$,
is axisymmetric with respect to the location of the star, and
the surface density scales as  $\Sigma(r) = \Sigma_0 \, r^{-1/2}$.
The material orbits initially on purely Keplerian orbits $v_r=0,
v_\varphi = G M_*/r^{1/2}$.
Deviations from Kepler rotation 
due to pressure gradients or self-gravity are typically
less than 1
thus this initial velocity setup is well justified. 
The fixed temperature law follows from the
constant vertical height $H/r = 0.05$ and is given by
$T(r) \propto r^{-1}$. The kinematic viscosity $\nu$ is given by an 
$\alpha$-description $\nu = \alpha c_s H$, with the sound speed $c_s$.
We present two models: 
\begin{enumerate}
\item[i)] 
Model A, having a constant $\alpha =0.01$, which may be on the high side
for protoplanetary disks but allows for a sufficiently rapid
evolution of the system to identify clearly the governing physical
effects. The two embedded planets have a mass of 3 and 5 $M_{Jup}$ and
are placed initially at 4 and 10 AU, respectively.
\item[ii)]
Model B with a constant $\nu$, equivalent to 
an $\alpha =0.004$ at 1 AU.
Here the two embedded planets each have an initial masses of 1 $M_{Jup}$, and
are placed initially at 1 and 2 AU, respectively.
This model is in fact a continuation of the 
one presented in \cite{2000MNRAS.313L..47K}.
\end{enumerate}
All the relevant model parameters are given in Table~\ref{tab:modpar}.
\subsection{Damped N-body}
\label{subsec:model-Nbody}
As pointed out, the full hydrodynamical evolution is computationally very
time consuming because ten-thousands of orbits have to be calculated.
Hence, we perform also pure N-body calculations
of a planetary system,
consisting only of a star and two planets.
The influence of the surrounding disk is felt here only through additional
(damping) forces.
For those we assume, that they act on the
semi-major axis $a$ and the eccentricity $e$ of the two planets
through an explicitly specified relations $\dot{a} (t)$ and $\dot{e} (t)$,
which may depend on time.
Here, as customary in solar system dynamics, it is assumed that
the motion of the two planets can be described at all times by approximate
Kepler ellipses where the time-dependent parameter
$a(t)$ and $e(t)$ represent the values
of the osculating orbital elements at the epoch $t$.

This change of the actual semi-major axis $a$ and the eccentricity $e$
caused by the gravitational action of the disk can be translated into
additional forces changing directly the position $\dot{\vec{x}}$
and velocity $\dot{\vec{u}}$ of the planets. In our implementation we
follow exactly \cite{2002ApJ...567..596L}, who give the detailed 
explicit expressions for these damping terms
in their appendix.
As a first test of the method we recalculated their model for
GJ~876 and obtained identical results.

Using the basic idea of two planets orbiting inside of a disk cavity
(see Fig.~\ref{fig:overview} below), we only damp $a$ and $e$ of
the outer planet.
Here we choose a general given functional dependence of the form
$g(t) = g_0 \cdot \exp[- (t/t_0)^p]$, with $g \in (a, e)$, 
where $a_0$ and $e_0$ are just the initial values.
The values of the exponent $p$ and the
timescale $t_0$ are adjusted to match the results of the full hydrodynamic
calculations. Comparative results of the two methods are displayed 
in Section \ref{subsec:Model-B}.
\section{Results}
\label{sec:results}
The basic evolutionary sequence of two planets evolving
simultaneously with the disk has been calculated and described by
\cite{2000MNRAS.313L..47K}  and \cite{2000ApJ...540.1091B}.
Before concentrating on details of the resonant evolution we first
summarize briefly the main results.
\begin{figure}[h]
\begin{center}
\resizebox{0.9\linewidth}{!}{%
\includegraphics{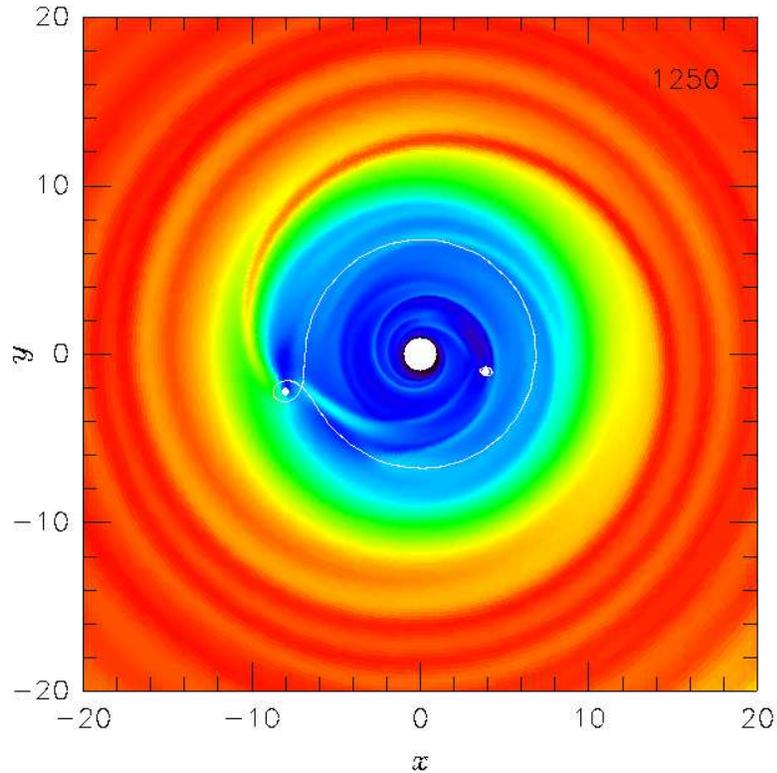}}
\end{center}
  \caption{
  Overview of the density distribution of model A
  after 1250 orbital periods of the inner planet.
  Higher density regions are brighter and lower ones are darker.
  The star lies at the center of the white inner region inside
  of $r_{min}=1$. The location of the two planets is indicated by the
  white dots, and the Roche-lobe sizes are also drawn. Clearly seen are
  the irregular spiral wakes generated by the planets. Only outside of the
  2nd planet the regular inter-twined two spiral arms are visible.
    }
   \label{fig:overview}
\end{figure}
\subsection{Full Hydrodynamic Evolution: Overview}
\label{subsec:fullhydro}
At the start of the simulations both planets are placed into
the axisymmetric disk, where the density is initialized such that
in addition to the radial density profile partially opened gaps are
superimposed.
Upon starting the evolution the two main effects are:
\begin{enumerate}
\item[a)]
As a consequence of the accretion of gas onto the two planets the
radial regime in between them will be depleted in mass and finally
cleared. This phase typically takes only a few hundred orbital periods. 
At the same time the region interior of the inner planet
will lose material due to accretion onto the central star.
Thus, after an initial transient phase we typically expect the
configuration of two planets
orbiting inside of an inner cavity of the disk,
see Fig.~\ref{fig:overview}, and also \cite{2000MNRAS.313L..47K}.
\item[b)]
After initialization the planets quickly (within a few orbital
periods) create non-axisymmetric disturbances, the spiral features,
in the disk. In contrast to the single planet case these are
no longer stationary in time, because there is no preferred rotating
frame.
The gravitational torques exerted on the two planets
by those density perturbations induce a migration process for the planets.
\end{enumerate}
\begin{figure}
\begin{center}
\resizebox{0.9\linewidth}{!}{%
\includegraphics{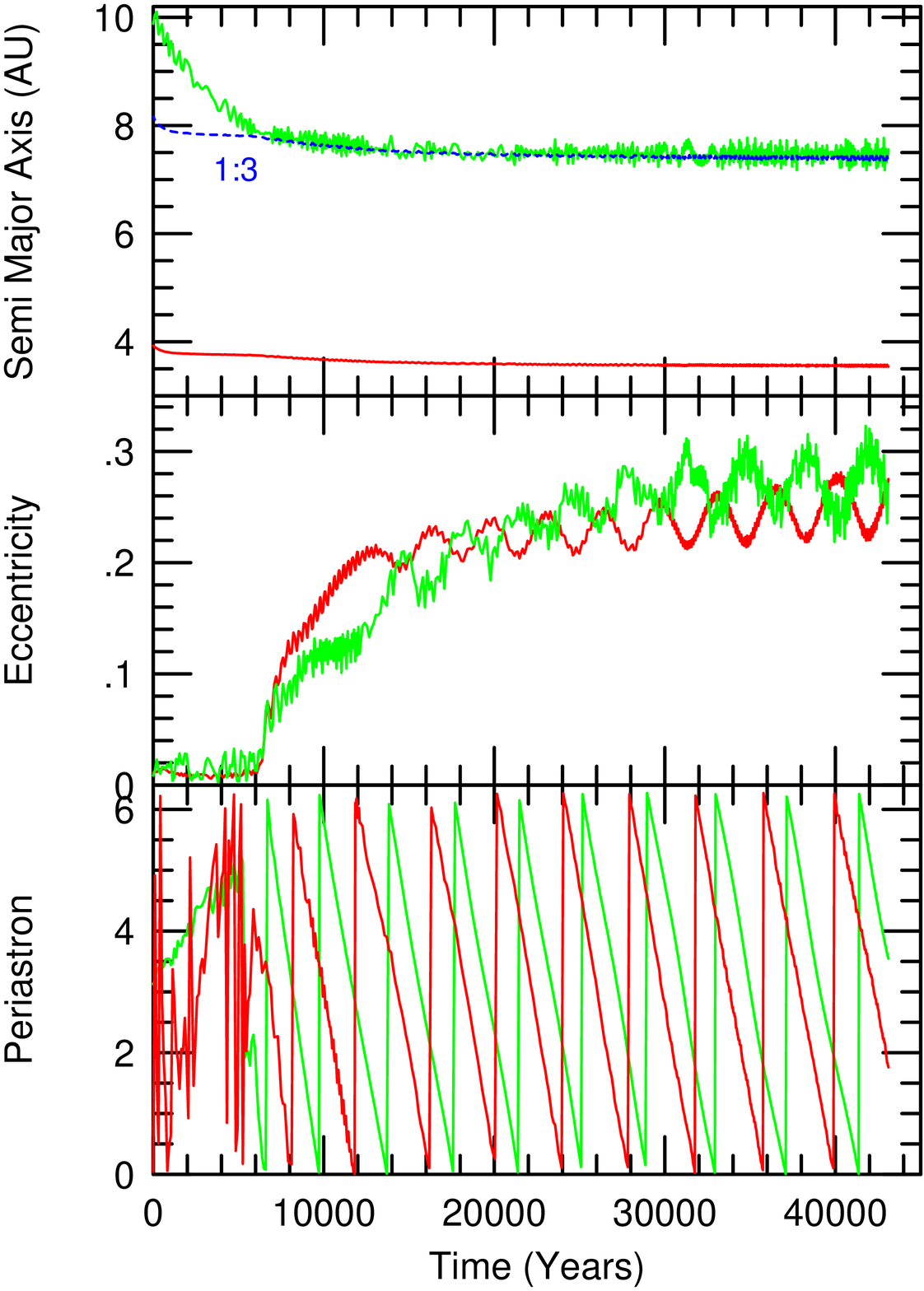}}
\end{center}
  \caption{The semi-major axis (AU), eccentricity and position of the
periastron of the orbit versus time for Model A.
In this example, the planets have fixed masses of
3 and 5 $M_{Jup}$, and are placed initially at 4 and 10 AU,
respectively. In the beginning, after the inner gap has cleared, only
the outer planet migrates inward, and the eccentricities
of both planets remain relatively small, less than $\approx 0.02$.
After about 6000 years the outer planet has reached a radius
with a period exactly three times that of the inner planet.
The periodic gravitational forcing leads to
a capture of the inner planet into a 3:1 resonance by the outer one.
This is indicated by the dark reference line (labeled 3:1), which
marks the location of the 3:1 resonance with respect to the inner planet.
Upon resonant capture the eccentricities grow, and the orbits librate
with a fixed relative orientation of $\Delta \omega = 110^o$.
}
\label{fig:aeo-tw6y}
\end{figure}
Now, the different radial location of the planets within the cavity
has a distinct influence on their subsequent evolution.
As a consequence of the clearing process the inner planet is no longer
surrounded by any disk material and thus cannot grow any further
in mass. In addition it cannot migrate anymore, because
there is no torquing material in its vicinity.
The outer planet on the other hand still has all the material
of the outer disk available, which exerts negative 
(Linblad) torques on the planet.
Hence, in the initial phase of the computations we observe an
inwardly migrating outer
planet and a stalled inner planet with a constant semi-major axis,
see the first 5000 yrs in the top panel of 
Fig.~\ref{fig:aeo-tw6y}.
 
This decrease in separation causes an increase of the gravitational
interaction between the two planets.
When the ratio of the orbital periods of the planets
has reached the fraction of two integers, i.e. they are in a
mean motion resonance, this
may lead to a resonant capture of the inner planet by the outer one.
Whether this happens or not depends on the physical conditions in the
disk (eq. viscosity) and the orbital parameter of the planets.
If the migration speed is too large for example, there may not be enough
time to excite the resonance, and the outer planet just continues its
migration process, see eg. \cite{1999MNRAS.304..185H}.
Also, if the initial eccentricities are too small, then
there may be no capture as well, see also \cite{2002ApJ...567..596L}.
\subsection{3:1 Resonance: Model A}
\label{subsec:Model-A}
In model A this capture
happens at $t \approx 6000$ when the outer
planet captures the inner one in a 3:1 resonance (see dark line
in top panel of Fig.~\ref{fig:aeo-tw6y}).
From that point on, the outer planet, which is still driven inward by the
outer disk material, will also be forcing the inner planet to
migrate inwards.
The typical time evolution of the orbital elements, semi-major axis ($a$),
eccentricity ($e$) and direction of the periastron ($\omega$),
of such a case are displayed in Fig.~\ref{fig:aeo-tw6y}
for model A.
\begin{figure*}
\begin{minipage}{0.48\linewidth}
\resizebox{\hsize}{!}{%
\includegraphics{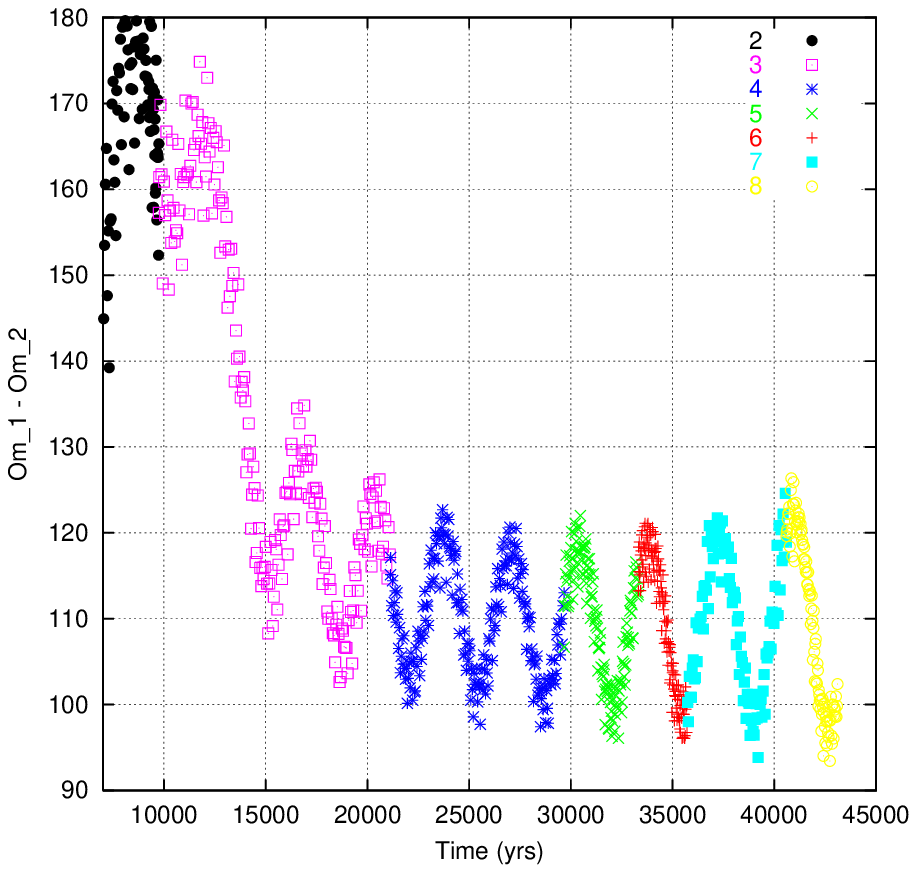}}
\end{minipage}
 \hfill \begin{minipage}{0.48\linewidth}
\resizebox{\hsize}{!}{%
\includegraphics{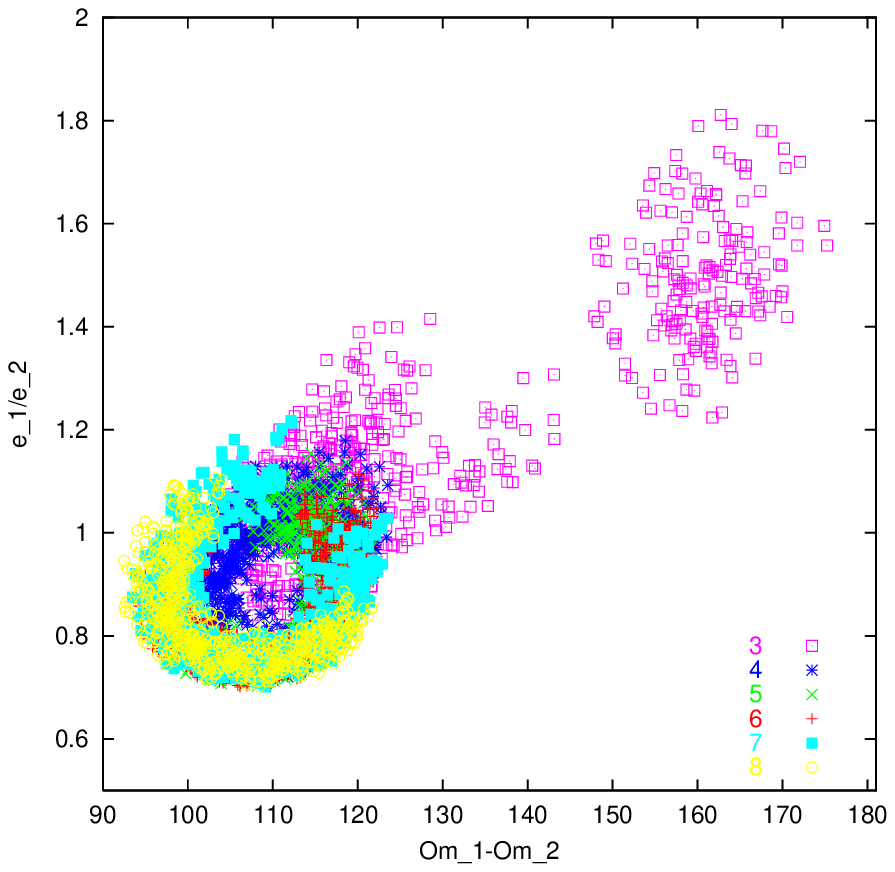}}
\end{minipage}
  \caption{Left: The difference in the direction of the periastron,
$\Delta \omega = \omega_1 - \omega_2$, of
the two planets vs. time. Right: Ratio of eccentricities $e_1/e_2$ versus
periastron difference. The indices 1,2 refer to the inner and outer planet,
respectively. The color and symbol coding is identical for the left and
right panel. During the evolution into resonance the dots are 'captured'
to eventually circle around the equilibrium value $\Delta \omega = 110^o$
and $e_1/e_2 = 0.9$.
}
\label{fig:o3-cor1.tw5}
\end{figure*}

We summarize the following important features of the evolution
after resonant capture:
\begin{enumerate}
\item[a)]
The inner planet begins to migrate inward as well, forced in by the outer
planet.
Thus both planets migrate inward simultaneously, always retaining their
resonant configuration.
As a consequence, the migration speed of the outer planet slows down,
and their radial separation declines.
\item[b)]
The eccentricities of both planets grow
initially very fast and then settle to oscillating
quasi-equilibrium values which change slowly on a secular time scale.
This slow increase of the eccentricities on the longer timescale
is caused by the growing gravitational forces between the planets,
due to the decreasing radial distance of the two planets
on their inward migration process.
\item[c)]
The ellipses (periastrae)
of the planets librate at a constant angular speed. Caused by the
resonance, the speed of libration for both planets is identical,
which can be inferred from the parallel lines
in the bottom panel of Fig.~\ref{fig:aeo-tw6y}. The orientation
of the orbits is phase locked with a constant separation
of the periastrae by a fixed phase-shift $\Delta \omega$.
\end{enumerate}
More detail of the capture into resonance and
the subsequent libration of the orbits
is illustrated in Fig.~\ref{fig:o3-cor1.tw5}
for model A. It is seen (left panel)
that the difference of the periastrae
settles to the fixed average value of
$\Delta \omega = 110^o$, a libration amplitude of about $15^o$, and 
libration period of about 3000 yrs.
The right panel shows the evolution in the $e_1/e_2$ vs $\Delta \omega$
using the same color and symbol coding coding.
During the initial process of capturing the points (open squares)
approach the final region from the top right region of the diagram.
At later times the points circle around the equilibrium point.
We note that additional models, not displayed here, with different
planet masses and viscosities all show approximately the
same shift in $| \Delta \omega |$, {\bf if} capture occurs
into 3:1 resonance.
The capture in such an asymmetric 3:1 resonance has been studied for the
55 Cnc case by \cite{astro-ph0209176}, and 
the stability of these resonant configurations has recently
been discussed by \cite{astro-ph0210577}.
\begin{figure*}
\begin{minipage}{0.48\linewidth}
\resizebox{\hsize}{!}{%
\includegraphics{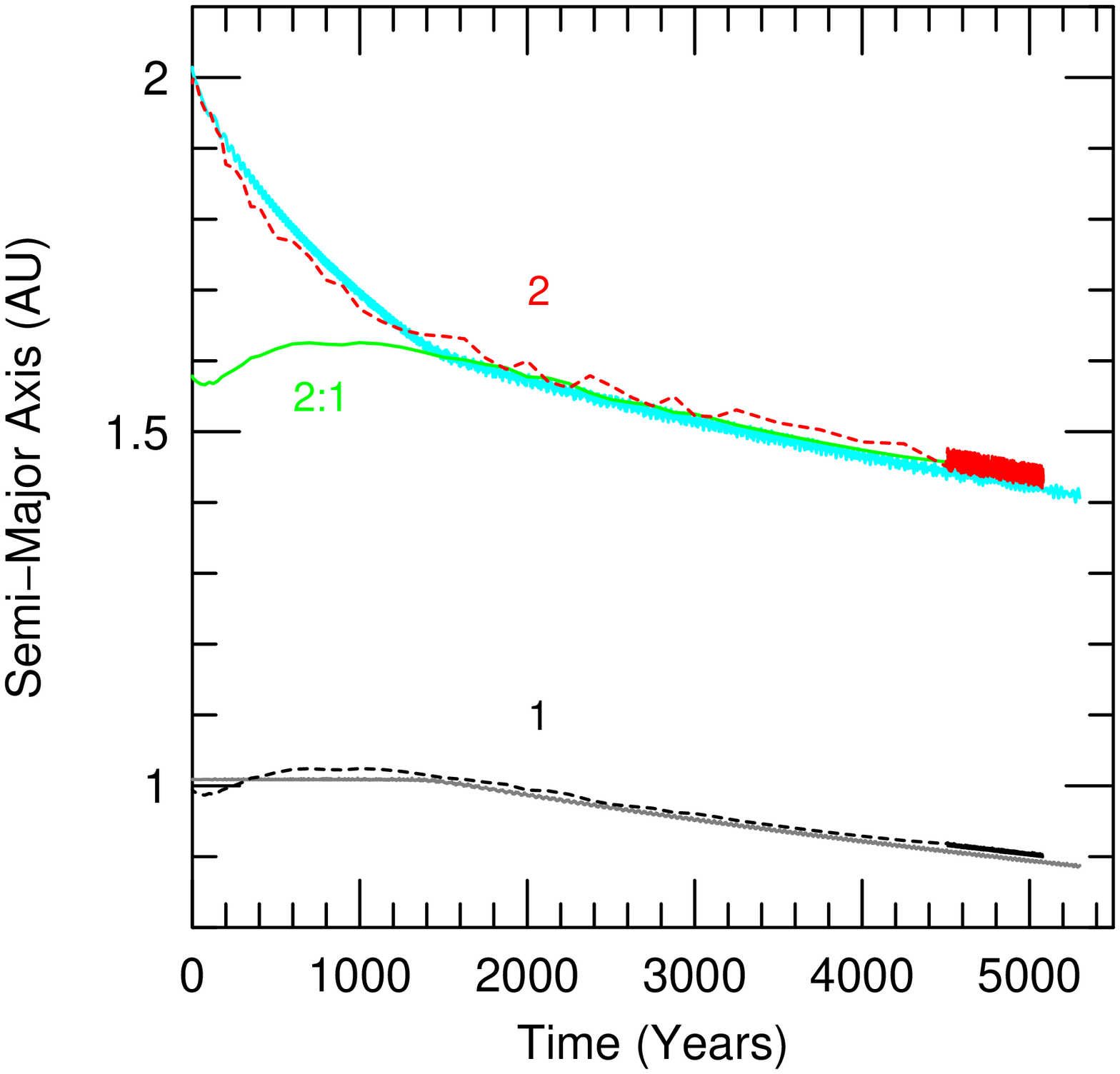}}
\end{minipage}
 \hfill \begin{minipage}{0.48\linewidth}
\resizebox{\hsize}{!}{%
\includegraphics{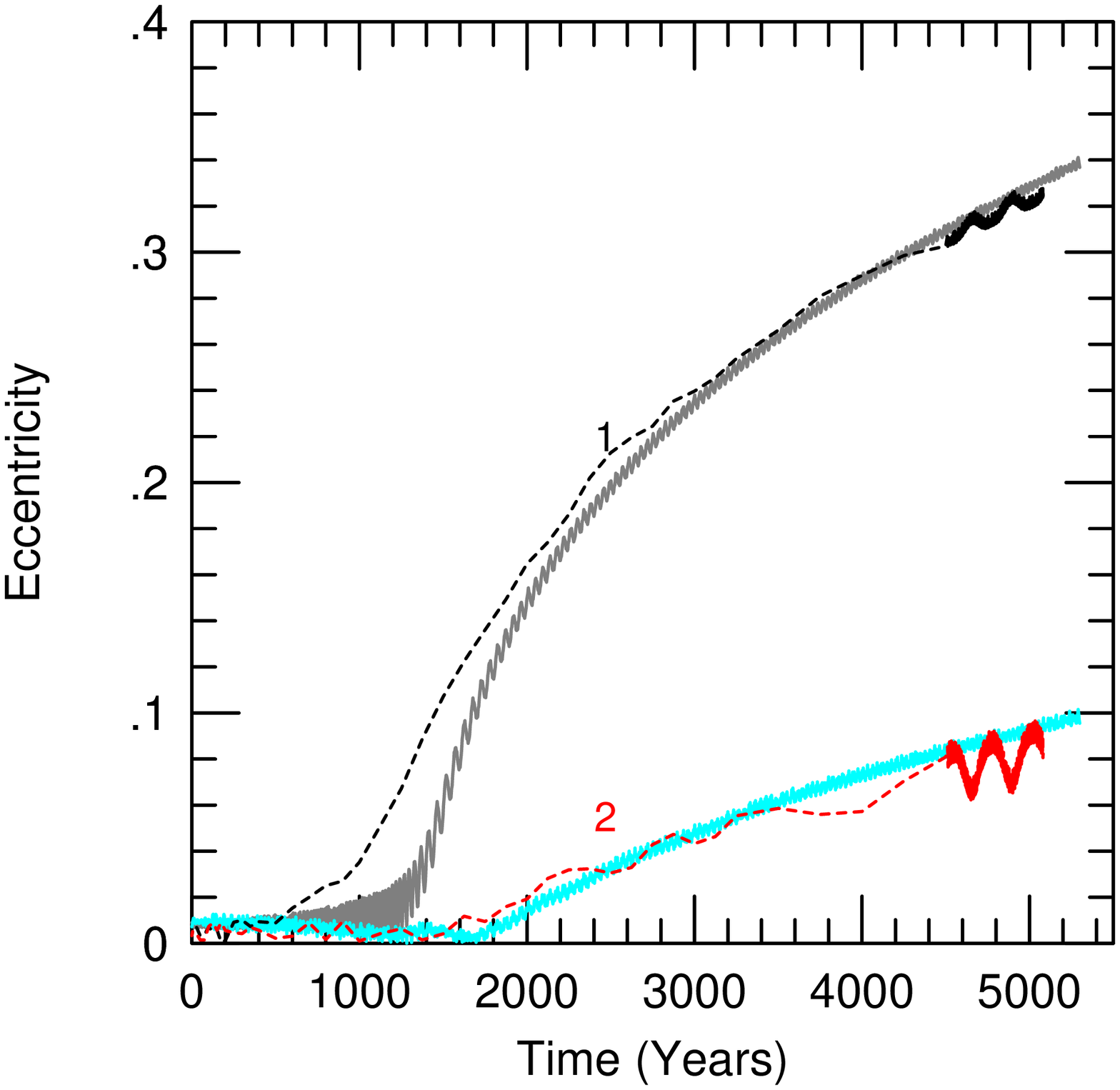}}
\end{minipage}
  \caption{The evolution of semi-major axis (left) and eccentricity
  (right) for model B. The planets had an initial radial location
   of 1 and 2 AU, and masses of 1 $M_{Jup}$ each, which were allowed
   to increase during the computation. 
  The results of the full hydrodynamic evolution
  are shown by the dashed lines. A reference line, with respect
  to the inner planet, indicating the location of the 2:1 resonance
  is shown. Before $t=4500$ only very few data points are plotted,
  thereafter they are spaced much more densely, which explains the different
  looking curves.
  The solid curves are obtained using the simplified damped
  3-body evolution as described in the text.
}
\label{fig:a-e.ps1b}
\end{figure*}
\begin{figure*}
\begin{minipage}{0.48\linewidth}
\resizebox{\hsize}{!}{%
\includegraphics{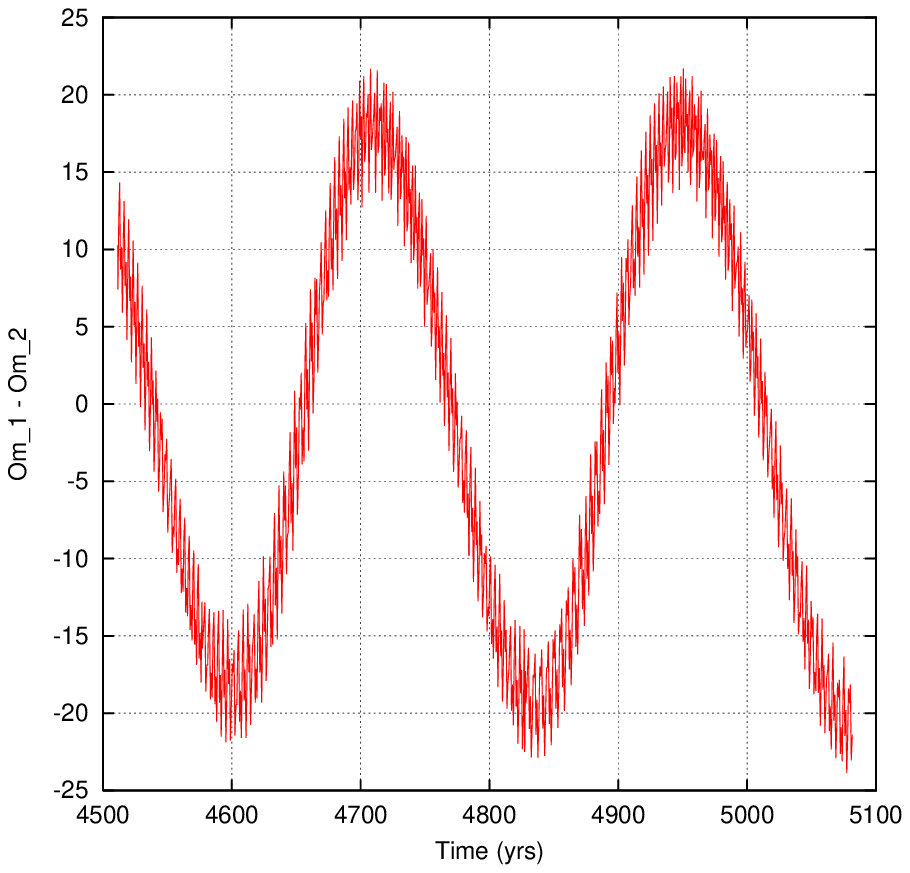}}
\end{minipage}
 \hfill \begin{minipage}{0.48\linewidth}
\resizebox{\hsize}{!}{%
\includegraphics{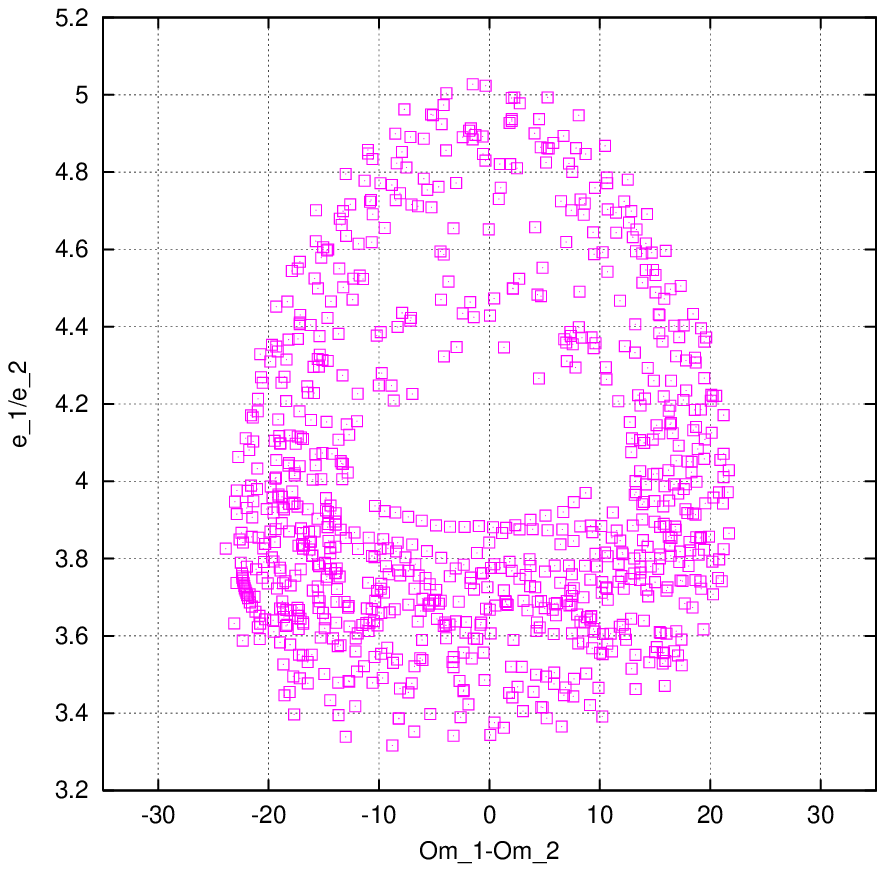}}
\end{minipage}
  \caption{Left: The difference in the direction of the periastrae of
the two planets vs. time. Right: Ratio of eccentricities versus
periastron difference. 
The data points are spaced equally in time with a distance of approximately
$\delta t = 3/4$ years.
Shown is only the very last section of the evolution
of model B, from 4500 to about 5100 yrs, which covers nearly two and a half
libration periods. In this case of a 2:1 resonance, the capture leads
to a complete alignment of the orbits with $\Delta \omega =0$.
}
\label{fig:o2-cor1.ps1b}
\end{figure*}
\subsection{2:1 Resonance: Model B}
\label{subsec:Model-B}
The second model setup is taken directly from 
\cite{2000MNRAS.313L..47K}. Here we continued exactly that model for
a little longer time, to infer some more characteristics of the 
intrinsic dynamics of that planetary system.
The evolution of the orbital elements $a$ and $e$ is displayed in
Fig.~\ref{fig:a-e.ps1b}. Here the planets are placed on initially
tighter orbits with a semi-major axis ratio of only 2. The initial orbital
evolution is similar to the first model, i.e. an inwardly migrating outer
and a stalled inner planet. However, caused by the reduced
initial radial distance higher resonances are not available,
and the resonant capture occurs into the
2:1 resonance, see reference line in left panel. The eccentricities
of both planets rise again upon capture but this time the mass
of the inner planet $m_1=1 M_{Jup}$ is,
due to the explained starvation, much smaller than
that of the outer one $m_2=3.1M_{Jup}$. This leads to 
to much larger rise in eccentricity, yielding a ration $e_1/e_2 \approx 4$.
In Fig.~\ref{fig:o2-cor1.ps1b} the alignment of the orbits is indicated.
This time, as seems typical for this type of 2:1 resonances
\cite{2001A&A...374.1092S,2002ApJ...567..596L}, the separation
in the periastrae is centered around zero, $\Delta \omega = 0$, with a
libration amplitude of up to about $20^o$.

For comparison and test, we modeled the evolution of the model B
also using the 3-body method, which is briefly outlined above
and compared this to the full hydrodynamic evolution.
As outlined in Sect.~\ref{subsec:model-Nbody}, only the outer
planet is damped in its semi-major axis $a$ and eccentricity $e$. It turned
out that for a good agreement the damping time scale $t_0$ is identical
for $a$ and $e$. 
Here in model B, we used $t_0 = 75,000$ yrs and $p=0.75$
to obtain the displayed results, and to reach the agreement
with the full hydrodynamical evolution. 
This contrasts the results of \cite{2002ApJ...567..596L} where a
shorter timescale was used for the eccentricity damping. The
difference may be caused for example by not modeled damping processes or
possible disk dissipation. In further models we plan to relate the
parameter $t_0$ and $p$ to physical quantities of disk such as viscosity,
temperature.

These additional results are also displayed in
Fig.~\ref{fig:a-e.ps1b} with the solid lines.
For the semi-major axis (left panel) the agreement is
very good indeed. The obtained eccentricities are in good agreement as
well. The only difference is the lack in eccentricity oscillations
in the simplified damped 3-body model, which may again be caused by a change
in eccentricity damping in the full model which is not modeled properly
in the simplified 3-body version.
\section{Summary and Conclusion}
\label{sub:summary}
We have performed full hydrodynamical calculations simulating the
joint evolution of a pair of protoplanets together with their surrounding 
protoplanetary disk, from which they originally formed.
The focus lies on massive planets in the range of a few Jupiter masses.
For the disk evolution we solve the Navier-Stokes equations, and the
motion of the planets is followed using a 4th order Runge-Kutta method,
considering their mutual interaction, the stellar and the disk's 
gravitational field.
These results were compared to simplified (damped) N-body computations,
where the gravitational influence of the disk is modeled through analytic
damping terms applied to the semi-major axis and eccentricity.
 
We find that both methods yield comparable results, if the damping
constants in the simplified models are adjusted properly.
These constants should be obtained from the full hydrodynamical evolution.

Two types of resonant situations are investigated. In the first case
the initial radial separation of the two planets was sufficiently far
that they were captured into a 3:1 resonance. In this case, the capture
leads to an orbit alignment of $\Delta \omega = 110^o$. In the
second case (model B) the capture is in 2:1 with  $\Delta \omega = 0^o$.
These difference in $\Delta \omega$ for varying types of resonances
seems to be a robust and a generic feature, which is supported by
the observations of all three resonating planets.
Additionally, we find that the inner planet preferably has lower 
mass caused by the inner disk gap, and finally the lower mass planet
should have a larger eccentricity. These findings are indeed seen in the
observed 2:1 resonant planetary systems GJ~876 and HD~82943.
\bibliographystyle{klunamed}
\bibliography{kley}
\end{article}
\end{document}